\documentclass[english,aps,prb,showpacs,amsmath,amssymb,lengthcheck]{revtex4}
\usepackage[T1]{fontenc}
\usepackage[latin9]{inputenc}
\usepackage{amstext}
\usepackage{graphicx}

\makeatletter

\@ifundefined{textcolor}{}
{%
 \definecolor{BLACK}{gray}{0}
 \definecolor{WHITE}{gray}{1}
 \definecolor{RED}{rgb}{1,0,0}
 \definecolor{GREEN}{rgb}{0,1,0}
 \definecolor{BLUE}{rgb}{0,0,1}
 \definecolor{CYAN}{cmyk}{1,0,0,0}
 \definecolor{MAGENTA}{cmyk}{0,1,0,0}
 \definecolor{YELLOW}{cmyk}{0,0,1,0}
 }

\makeatletter

\usepackage{epsfig} 

\usepackage{hyperref}

\makeatother

\makeatother

\usepackage{babel}

\begin{document}
\global\long\def\rs{\rho_{s}}
 \global\long\def\ie{i.~e., }
 \global\long\def\bra#1{\left|#1\right>}
 \global\long\def\ket#1{\left<#1\right|}
 \global\long\def\comment#1{\emph{#1}}
 \global\long\def\likeatop#1#2{\genfrac{}{}{0pt}{}{#1}{#2}}
 \global\long\def\fs{\phi_{0}}
 \global\long\def\ofs{\overline{\phi}_{0}}
 \global\long\def\ff{\zeta}
 \global\long\def\ratio{\gamma}
 \global\long\def\fa{ \overline{\theta}}

\pacs{11.10.Gh, 11.10.Kk, 64.70.Rh}

\title{Pokrovsky-Talapov Model at finite temperature: a renormalization-group
analysis}

\author{{\small A. Lazarides, O. Tieleman and C. Morais Smith}}

\address{Institute for Theoretical Physics, Utrecht University,\\
 Leuvenlaan 4, 3584 CE Utrecht, The Netherlands.}

\date{\textit{\small \today}}
\begin{abstract}
We calculate the finite-temperature shift of the critical wavevector
$Q_{c}$ of the Pokrovsky-Talapov model using a renormalization-group
analysis. Separating the Hamiltonian into a part that is renormalized
and one that is not, we obtain the flow equations for the stiffness
and an arbitrary potential. We then specialize to the case of a cosine
potential, and compare our results to well-known results for the sine-Gordon model, to which our model reduces in the limit of vanishing driving wavevector $Q=0$. Our results may be applied to describe the commensurate-incommensurate phase transition in several physical systems and allow for a more realistic comparison with experiments, which are always carried out at a finite temperature.
\end{abstract}
\maketitle

\section{Introduction}

The Pokrovsky-Talapov\cite{bak,PTPRL} (PT) model describes a large
variety of systems displaying a commensurate/incommensurate (C/IC)
transition, ranging from vortex depinning in type-II superconductors~\cite{drose}
to adsorbate layers on crystal surfaces~\cite{COPd100} and quantum
Hall bilayer systems under a tilted magnetic field.~\cite{unification,bilayersmcd,hanna}
This model is closely related to the sine-Gordon model,\cite{sft}
with the extra feature of a characteristic driving wavevector $\mathbf{Q}$
imposed through the cosine term. In this work we present a functional
renormalization group calculation of the finite-temperature corrections
to the mean field results.

The Hamiltonian for the PT model is \begin{equation}
H_{PT}=\int d^{2}r\,\left[\frac{1}{2}\rs\left(\nabla\phi\right)^{2}-t\cos\left(\phi-Qx\right)\right],\label{PTorig}\end{equation}
 where $\mathbf{r}=(x,y)$ and $d^{2}r=dxdy$. At mean-field level,
one approximates the thermodynamic free energy by the Hamiltonian
itself, evaluated at the field configuration that minimises the energy
for given boundary conditions. This neglects the contributions of
all fluctuations away from the minimum, therefore becoming less accurate
as the temperature is raised and the entropic contribution of fluctuations
increases. In this approximation, it is straightforward to compare
the free energy of a configuration with the field following the driving
wavevector (commensurate phase) to that of one in which the field
no longer follows the imposed $Q$. As $Q$ increases, the presence
of the stiffness term, suppressing deviations of the field from uniformity,
makes it more and more costly to remain in the commensurate phase,
until a critical $Q_{c}$ is reached at which the incommensurate phase
becomes energetically favourable. The aim of this article is to investigate
the effects of thermal fluctutations on this critical $Q_{c}$.

We shall use a functional renormalization group (RG) scheme to study
the model at finite temperatures. Our approach is as follows: We perform
a simple transformation which maps the PT model to a sine-Gordon model
with additional terms depending only on the total topological {}``charge''
of the system and on the driving wavevector $Q$. The RG transformation
does not couple the sine-Gordon part to the $Q$-dependent part. Taking
advantage of this, we renormalize the sine-Gordon part of the Hamiltonian
and obtain a long-wavelength effective action, which we subsequently
use to obtain the new value of the critical $Q_{c}$.

The main technical complication one faces in constructing a systematic
renormalization group transformation for the sine-Gordon model is
the inability to expand the cosine term in Eq.~\eqref{PTorig} in
powers of the field and keep only a finite number of these. There
are three main reasons for this complication: First, the periodicity
of the cosine is crucial and is destroyed by any finite-order Taylor
expansion. Second, we are interested in the two-dimensional case;
as follows from simple power counting,\cite{amit,sft} polynomial
interactions involving any power of the field are relevant in two
dimensions. Finally, we are not interested in the fixed point of the
RG transformation, but in the actual values of the various parameters
after integrating out the degrees of freedom that we are not interested
in.

Not expanding the cosine means that, in diagrammatic language, we
ought to keep track of an infinite number of diagrams to one loop.
This problem has already been solved for the case of the Wilsonian
RG by F. Wegner and A. Houghton in Ref.~\onlinecite{wegnerhoughton},
where these authors derive the eponymous exact renormalization group
equation. This equation is also the limit of an approximate recursion
relation first given by Wilson.\cite{wilsonkogut} Wilson's approximate
recursion relation has been applied to the problem of critical wetting
in Refs.~{[}\onlinecite{nlrglipfish,wettingboulter}{]} because
the effective field theories used in studying critical wetting share
with our problem its dimensionality, preventing the use of the more
usual perturbation methods. In general, functional renormalization
group approaches are useful for cases where there is an effective
potential with a nontrivial functional dependence on the field, such
as the cosine term in Eq. \eqref{PTorig}.

The exact functional renormalization equation of Wegner and Houghton
relies on a sharp, moving cutoff in momentum space. A sharp cutoff
induces long-range interactions in real space and complicates the
calculation of the flow of the stiffness $\rs$ in Eq. \eqref{PTorig}.
This may be overcome by employing a smooth cutoff function;\cite{polchinski,huse}
however, the resulting trajectories depend on the precise form of
the cutoff. It has been shown\cite{schemedependence} that the Wilsonian
RG approach (of which the Polchinksi RG is an example) suffers from
strong scheme dependence even in the asymptotic regime. Various alternative
formulations of RG transformations exist that do not suffer from this
problem; one that has recently been applied to the sine-Gordon model
is the functional renormalization of the effective average action\cite{wetterich,polonyisinegordon}
(EAARG) in which a transformation is obtained, not for the Hamiltonian
itself, but for the generating function for the 1-particle irreducible
Green's functions. This is the RG scheme that we use in this article.

The outline of this paper is as follows: In Sec.~\ref{secmodel}
we describe the model and give a qualitative description of its behaviour.
In Sec.~\ref{secdeterminationofQc} we explain the basic idea behind
our approach before proceeding directly to the derivation (Sec.~\ref{secdervfloweqns})
and application to the PT model (Sec.~\ref{secapplrgpt}) of the
appropriate RG flow equations. Finally, we give a brief discussion
of our results in Sec.~\ref{secconclusions}.

\section{The Model}

\label{secmodel} To qualitatively understand the features exhibited
by a system described by the PT Hamiltonian given in Eq.~\eqref{PTorig},
consider first the case $Q=0$. It is then clear that, at mean field
level, $\phi$ will simply remain at one of the minima of the potential
$V(\phi)=-t\cos(\phi)$ at $\phi_{n}=2\pi n$ with $n=0,\pm1,\pm2,\ldots$

On the other hand, consider the quantity $\left<\phi(\mathbf{r})\phi(\mathbf{r}')\right>_{c}=\left<\phi(\mathbf{r})\phi(\mathbf{r}')\right>-\left<\phi(\mathbf{r})\right>\left<\phi(\mathbf{r}')\right>$;
for low enough temperatures, it is given by the 1-loop result \begin{equation}
\left<\phi(\mathbf{r})\phi(\mathbf{r}')\right>_{c}=\frac{k_{B}T}{\rho_{s}}K_{0}\left(\frac{|\mathbf{r}-\mathbf{r}'|}{\xi}\right)\label{eq1loopcorr}\end{equation}
 in which $\xi=\sqrt{\rs/t}$, $K_{0}(x)$ is a modified Bessel function
and $\left<\ldots\right>_{c}$ indicates a cumulant. Since $\xi\rightarrow\infty$
if $t\rightarrow0$ and $K_{0}(x)=\ln(2/x)+\mathrm{const}+O(x^{2}\ln(x))$,
the right hand side of Eq.~\eqref{eq1loopcorr} diverges as $t\rightarrow0$,
\ie there is no long-range order in the system. This is merely an
example of the Mermin-Wagner theorem.\cite{sft,chaikinlubensky} In
addition, the $Q=0$ system exhibits a Kosterlitz-Thouless type transition\cite{chaikinlubensky}
on some line $t(\rs)$, which is again completely missed by a mean-field
analysis.

This transition is analogous to the roughening transition in interface
problems.\cite{huse,chuiweeks} In this analogy, the phase $\phi$
corresponds to the height of an interface above a reference level.%
\footnote{That is, a noncompact variable; some complications from the
compactness of $\phi$ are taken into account in
Ref.~[\onlinecite{fertig}].} The phase in which
$\left<\phi(\mathbf{r})\phi(\mathbf{r}')\right>_{c}$ remains finite as
$|\mathbf{r}- \mathbf{r}'|\rightarrow\infty$ is called the {}``smooth''
phase, while the one in which
$\left<\phi(\mathbf{r})\phi(\mathbf{r}')\right>_{c}$ diverges is called
the {}``rough'' phase. Given the value of $\phi(0)=\fs$ at some
arbitrary point which we take to be the origin, the value of
$\phi(\mathbf{r})$ at some other point $\mathbf{r}'$ arbitrarily far
from it either remains within a finite distance from $\phi_{0}$ (smooth
phase) or it does not but rather crosses over the maxima of $V(\phi)$.
Clearly, the system with $t=0$ is in the rough phase;\footnote{Note that
the $t=0$ system also undergoes a KT transition between a phase with
exponentially-decaying to one with algebraically-decaying correlations;
this transition is different from the one we mention here and will not
concern us in this work.} the usual RG analysis of the sine-Gordon
model, Eq.~\eqref{PTorig} with $Q=0$ shows that, for given temperature
$T$ and $\rs/T<1/8\pi$, there exists a $t_{c}$ below which the system is
rough and above which it is smooth. For $\rs/T>1/8\pi$ it is always
smooth.

Consider now the case of finite $Q$. In the roughening picture, this
corresponds to a potential $V$ that depends on the position $x$;
as $x$ increases, the minima of the potential move to larger values
of $\phi_{n}=2\pi n+Qx$. In other words, the potential is effectively
{}``tilted''. Thus, the potential part of the Hamiltonian tends
to favour a $\phi$ that increases with position and follows the potential,
$\phi=\phi_{n}$ ({}``commensurate phase''), while the gradient
part favours a spatially constant $\phi$ ({}``incommensurate phase'').
The competition between $\rs$ and $t$ leads to a transition between
the two states as, for example, $t$ is varied.

Notice that there are two separate effects here: one is the roughening
transition (belonging to the Kosterlitz-Thouless universality class),
which is already present when $Q=0$ and the other is the commensurate-incommensurate
transition, which appears only for finite $Q$. These two effects
may be conveniently separated out as described in the next section.

\section{Determination of $Q_{c}$ at mean-field level}

\label{secdeterminationofQc}

\subsection{Separation of the Hamiltonian}

\label{subsecseparationoftheHamiltonian} We begin by shifting to
the new variable $\theta=\phi-Qx$, whereupon the Hamiltonian becomes
\begin{subequations} \label{eqHamiltonianall} \begin{equation}
H_{PT}[\theta]=H_{sG}+H_{Q},\label{PTtransformed}\end{equation}
 with \begin{equation}
H_{sG}[\theta]=\int d^{2}r\,\left[\frac{1}{2}\rs\left(\nabla\theta\right)^{2}-t\cos\left(\theta\right)\right],\label{eqHsgdefinition}\end{equation}
 \begin{equation}
H_{Q}(n_{T}^{\theta})=\frac{1}{2}\rs Q^{2}+2\pi\rs Qn_{T}^{\theta}\label{eqHqdefinition}\end{equation}
 and \begin{equation}
n_{T}^{\theta}=\frac{1}{2\pi}\int dx\,\partial_{x}\theta=\sum_{i}n_{i}^{\theta};\label{eqcharge}\end{equation}
 \end{subequations} 
that is, the sum of the {}``charges'' of all solitons present (per
unit length); the integer $i$ simply indexes the soliton.

This form of the Hamiltonian is advantageous in that it consists of
a sine-Gordon part, $H_{sG}$, which is independent of $Q$, plus
the two terms in $H_{Q}$, which do depend on $Q$. This last term
is the essential difference from a simple sine-Gordon model. As we
shall see in Sec.~\ref{secdervfloweqns}, $H_{Q}$ is unaffected
by an RG transformation. This will form the basis of our treatment
of finite temperature effects.

\subsection{Determination of critical $Q_{c}$}

\label{sec:mean_field_determination_of_critical_q_}

We shall next compute the critical $Q_{c}$ in the mean-field approximation;
this will be straightforwardly extended to the renormalized case in
Sec.~\ref{secapplrgpt}.

To obtain the critical $Q_{c}$, we notice that the transition from
the commensurate to the incommensurate phase is signalled by the appearance
of a finite soliton density with charge $-|n_{T}^{\theta}|$. We therefore
calculate the excess energy per unit area of a configuration with
a single soliton (with charge $n^{\theta}=-1$) over that of one with
no solitons, $\theta=0$; we denote this by $\Delta E$. The part
of the energy cost of a solitonic configuration due to $H_{sG}$ is
calculated in a standard way (sketched out in Appendix~\ref{appsolitonenergy})
to be $E_{\mathrm{sol}}=8\sqrt{\rs t}$, hence $\Delta E=E_{\mathrm{sol}}-2\pi\rs Q$.
This quantity vanishes at the transition point, yielding a mean-field
critical $Q_{c}$ of~\cite{hanna} \begin{equation}
Q_{c}=\frac{4}{\pi}\sqrt{\frac{t}{\rs}}.\label{eqcriticalQc}\end{equation}
Note that $Q_{c}$ diverges as $\rs\rightarrow0$, implying that
the system remains in the commensurate state for all momenta; this
is in agreement with the discussion at the end of Sec.~\ref{secmodel},
according to which the creation of solitons (hence the transition
to the incommensurate state) is caused by the stiffness overcoming
the tendency of the phase field $\phi$ to follow the minimum of the
tilted potential.

\section{Exact renormalization group equations}

\label{secdervfloweqns}

Various schemes have been developed to study renormalization group
transformations of two dimensional field theories: In the theory of
critical wetting, there have been several studies using this formulation
of the RG, initially to first order in the potential~\cite{frgfishhuse}
and then to all orders.~\cite{nlrglipfish} All these approaches have
essentially used a local-potential approximation (LPA), in which the
potential is allowed to change under coarse-graining, while the
stiffness is not. The same method was later extended to a nonlocal
model.~\cite{nlwettingPRL} In the same context there has also been work
in which the gradient term is renormalized in an approximate
way.~\cite{wettingboulter} The LPA is generally thought to be applicable
to wetting problems because the anomalous dimension is expected to be
zero.

In the present problem, on the other hand, it is clear that for vanishing
$Q$ one should obtain Kosterlitz-Thouless behaviour; furthermore,
as discussed above, the quantity $Q$ only couples to the number of
solitons, which is conserved under the RG flow. Thus, the appropriate
RG must capture the Kosterlitz-Thouless type of behaviour, which requires
$\rs$ to flow under the transformation.

As mentioned above, we shall use the effective average functional
RG scheme introduced by Wetterich\cite{wetterich} and applied recently
to the sine-Gordon model.\cite{polonyisinegordon} For completeness,
we outline the derivation of the exact flow equation for the effective
action before applying it to the sine-Gordon model; for more details,
see Refs.~\onlinecite{wetterich,alexandreannphys,polonyisinegordon}.

From this point onwards, we will subsume the temperature into the
parameters $\rs$ and $t$; that is, we use units in which the temperature
$k_{B}T=1$.

One begins by defining the bare action $\mathcal{S}[\theta]=H[\theta]$
and adding to it a piece \[
\Delta\mathcal{S}_{R}[\theta]=\frac{1}{2}\sum_{\mathbf{q}}R_{\Lambda}(\mathbf{q)\theta(\mathbf{q})\theta^{*}(-\mathbf{q})}\]
where $R_{\Lambda}(\mathbf{q})$ is called a regulator function (see
below). One also adds a source term $\Delta\mathcal{S}_{S}[j,\theta]=\Sigma_{q}\left(j(\mathbf{q)}\theta^{*}(-\mathbf{q})+j^{*}(-\mathbf{q)}\theta(\mathbf{q})\right)$
and writes $\mathcal{S}_{\Lambda}[j,\theta]=\mathcal{S}[\theta]+\Delta\mathcal{S}_{R}[\theta]+\Delta\mathcal{S}_{S}[j,\theta]$.
The quantity \[
W_{\Lambda}[j]=\log\int\mathcal{D}\theta\exp\left(-S_{\Lambda}[j,\theta]\right)\]
 is then the generator of connected correlation functions\cite{amit}
for the action $\mathcal{S}_{\Lambda}$. Its Legendre transform is
\[
\tilde{\Gamma}_{\Lambda}[\fa]+W_{\Lambda}[j]=\Sigma_{\mathbf{q}}\left(j^{*}(-\mathbf{q})\fa(\mathbf{q})+j(\mathbf{q})\fa^{*}(-\mathbf{q})\right),\]
 where $\fa(\mathbf{q})=\delta W[j]/\delta j^{*}(-\mathbf{q})$ is
the average of the field. We also define a new, related function \begin{equation}
\Gamma_{\Lambda}[\fa]=\tilde{\Gamma}_{\Lambda}[\fa]-\Delta\mathcal{S}_{R}[\fa].\label{eqGammadefn}\end{equation}

Using well-known\cite{amit} properties of $W$ and $\Gamma$, and
writing $\epsilon=\ln(\Lambda_{0}/\Lambda)$ where $\Lambda_{0}$
is the initial value of the cutoff $\Lambda$, one finds \begin{equation}
\partial_{\epsilon}\Gamma_{\Lambda}[\fa]|_{\fa}=\frac{1}{2}\text{Tr}\left(\partial_{\epsilon}R_{\Lambda}(\mathbf{q})\left[\Gamma_{\Lambda}^{(2)}[\fa]+R_{\Lambda}(\mathbf{q})\right]^{-1}\right)\label{eqwetterichflow}\end{equation}
with $\Gamma_{\Lambda}^{(n)}(\fa)$ indicating the $n^{th}$ functional
derivative of $\Gamma_{\Lambda}$. This is an exact result.\cite{wetterich}
It can be shown\cite{wetterich} that if $R_{\Lambda}(q)\to\infty$
as $\Lambda\to\infty$ then $\Gamma_{\Lambda}[\fa]\to S_{\Lambda}[\fa]$:
fluctuations about the mean-field solution are completely suppressed.
Conversely, if $R_{\Lambda}(q)\to0$ as $\Lambda\to0$ then $\Gamma_{\Lambda}[\fa]\to\Gamma[\fa]$,
so that the full generator of 1 particle irreducible (1PI) vertices
is obtained.

A full solution of Eq.~\eqref{eqwetterichflow} for $\Gamma_{\Lambda}$
would amount to computing all 1PI functions of the system at some
length scale $\sim1/\Lambda$, including the full effects of fluctuations.
This is not a simple problem, and one must resort to approximations.
We will take the form of $\Gamma_{\Lambda}[\fa]$ to be \[
\Gamma_{\Lambda}[\fa]=\int d^{2}r\left(\frac{1}{2}\rho_{s}(\epsilon)(\nabla\fa)^{2}+V(\epsilon,\fa)\right).\]
 A tedious but straightforward computation leads to\cite{bonanno,Branchina,alexandreannphys}
\begin{widetext}\begin{subequations} 
\begin{align}
\partial_{\epsilon}V= & 2V-\frac{1}{2}\int\frac{d^{2}q}{(2\pi)^{2}}\mathcal{A}(\mathbf{q})\partial_{\epsilon}R_{\Lambda}(\mathbf{q})\label{eqflowVgeneral}\\
\partial_{\epsilon}\rs= & \frac{1}{2}\mathcal{P}V^{(3)}\int\frac{d^{2}q}{(2\pi)^{2}}\mathcal{A}^{4}(\mathbf{q})\partial_{\epsilon}R_{\Lambda}(\mathbf{q})\left(-2\rs(\epsilon)+\rs^{2}(\epsilon)\mathcal{A}(\mathbf{q})\mathbf{q}^{2}\right)\label{eqflowrsgeneral}\
\end{align}
 \end{subequations}\end{widetext}where the operator $\mathcal{P}$
projects the function to its right onto the field-independent functional
subspace%
\footnote{That is, $\mathcal{P}[\cdot]=\int_{0}^{2\pi}d\fa[\cdot]/2\pi$.%
} and $\mathcal{A}(\mathbf{q})=(\rs q^{2}+R_{\Lambda}(\mathbf{q})+V^{(2)})^{-1}$.
We take the cutoff function to be \begin{equation}
R_{\Lambda}(q)=q^{2}\left(\frac{\Lambda^{2}}{q^{2}}\right)^{b}.\label{eqregulator}\end{equation}
The parameter $b$ controls the sharpness of the regulator function
$R_\Lambda$ in both wavevector and real space: for large $b$, $R_\Lambda$ is local in wavevector space and long-range in real space; for
$b\rightarrow 1$, it is instead smooth in momentum space but sharp in real space. 

If $V(\epsilon,\fa)$ is restricted to its leading Fourier component
$V(\epsilon,\fa)=-t(\epsilon)\cos(\fa)$, the flow equations are,
to leading-order and after rescaling, \begin{eqnarray*}
\partial_{\epsilon}t(\epsilon) & = & \left(2-\frac{1}{4\pi\rs}\right)t(\epsilon)/\Lambda^{2}\\
\partial_{\epsilon}\rs(\epsilon) & = & \frac{\left(t(\epsilon)/\Lambda^{2}\right)^{2}}{\left(\rs(\epsilon)\right)^{2-2/b}}\tau_{b}\end{eqnarray*}
with $\tau_{b}=b\Gamma(3-2/b)\Gamma(1+1/b)/(48\pi)$, reproducing
the well-known leading-order flow equations for the sine-Gordon model.\cite{josekadanoffkirkpatricknelson,huse,knopsouden,sft} 

\begin{figure}[t]
\centering \includegraphics[width=8cm]{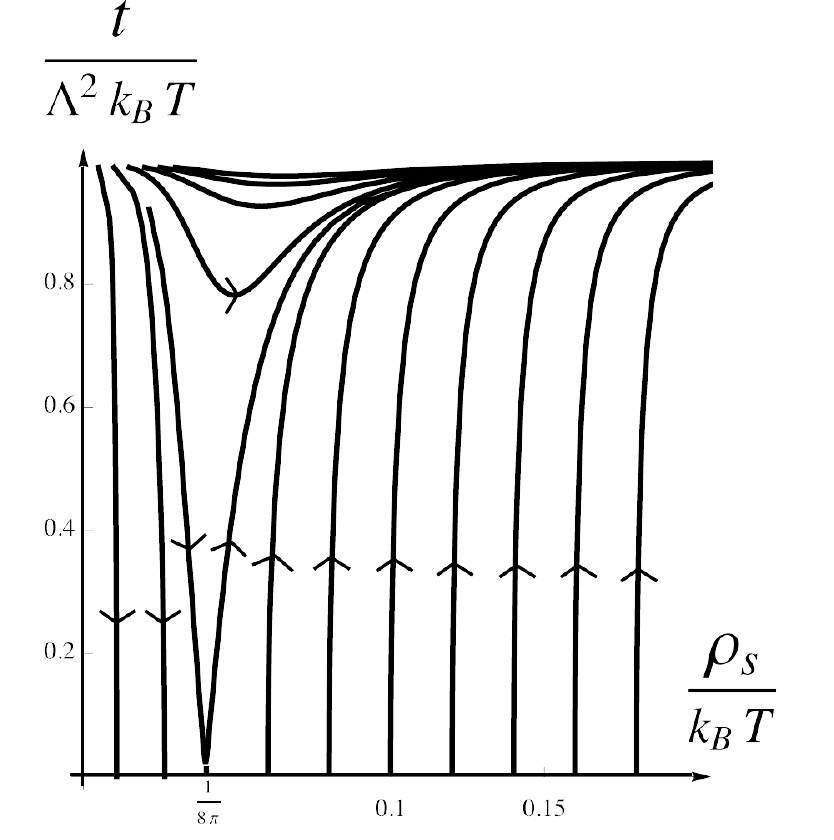} 

\caption{Flow diagrams for the RG Eqs.~\eqref{eqflowtrestricted} and~\eqref{eqflowrsrestricted}. }

\label{figflowdiag} 
\end{figure}

Including higher-order terms on the right hand side of
Eqs.~\eqref{eqflowVgeneral} and~\eqref{eqflowrsgeneral} is, in principle,
straightforward; for example, expanding $\mathcal{A}(\mathbf{q})$
in powers of $V(\epsilon,\fa)$ and computing the integrals, one obtains
\begin{widetext}
\begin{equation}\begin{split}
\partial_\epsilon V(\epsilon,\fa)=
-\frac{V^{(2)}}{2 b \Lambda  \rho } +\frac{(-1+b) \pi   \left(\frac{\rho }{\Lambda }\right)^{1+\frac{1}{b}}(V^{(2)})^2} {4 b^3 \rho ^3\sin \left(\frac{2\pi}{b}\right)} -\frac{(-2+b) (-1+b) \pi \left(\frac{\rho }{\Lambda }\right)^{\frac{2+b}{b}}   (V^{(2)})^3}{3 b^4 \rho ^4\sin \left(\frac{2\pi}{b}\right)} +\ldots
\end{split} \label{eqseries}\end{equation}
\end{widetext}
A similarly ungainly equation holds for $\partial_{\epsilon}\rs(\epsilon)$.
These expressions are greatly simplified if we make the choice $b=1$
in the regulator function Eq.~\eqref{eqregulator}. For the potential,
the series on the right hand side of Eq.~\eqref{eqseries} may be
summed (see Appendix~\ref{apptrigident}) to yield\[
\partial_{\epsilon}V=\frac{\Lambda}{4\pi\rs}\ln\left(1+\frac{V^{(2)}}{\Lambda^{2}}\right).\]
Restricting the potential to the form
$V(\epsilon,\fa)=-t(\epsilon)\cos(\fa)$ we obtain the flow equation for
$t(\epsilon)$\cite{polonyisinegordon}
\begin{equation}
\partial_{\epsilon}t(\epsilon)=2t(\epsilon)-\frac{1}{2\pi\rs(\epsilon)t(\epsilon)/\Lambda^{2}}\left(1-\sqrt{1-\left(\frac{t(\epsilon)}{\Lambda^{2}}\right)^{2}}\right)\label{eqflowtrestricted}\end{equation}
(see Appendix~\ref{apptrigident}). In a similar way, and using the results of Appendix~\ref{apptrigident} to isolate the field-independent part, we obtain for the flow of $\rs(\epsilon)$
\begin{equation}
\partial_{\epsilon}\rs(\epsilon)=\frac{t^{2}(\epsilon)/\Lambda^{2}}{24\pi\left(1-t^{2}(\epsilon)/\Lambda^{4}\right)^{3/2}}.\label{eqflowrsrestricted}\end{equation}
The flow diagram corresponding to Eqs.~\eqref{eqflowtrestricted} and
\eqref{eqflowrsrestricted} is shown in Fig.~\eqref{figflowdiag}.

We now turn to the application of the flow equations to the PT model.

\section{Application of RG to the PT model}

\label{secapplrgpt}

\subsection{Calculation of $Q_{c}$ using the RG results}

\label{secdeterminationQcRG}

\begin{figure}[t]
\centering \includegraphics[width=8cm]{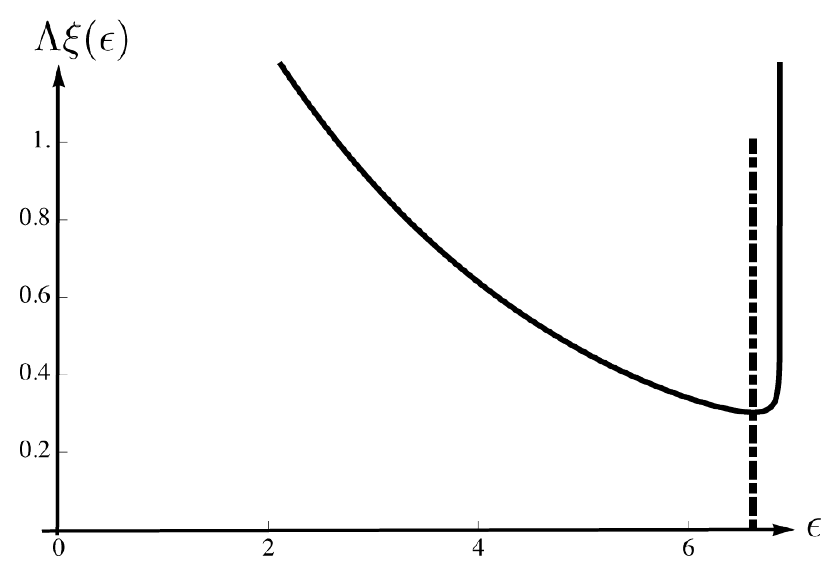} 

\caption{Evolution of $\Lambda\xi(\epsilon)$ (dark full line) as a function
of $\epsilon$. The dot-dashed line indicates the position of the
minimum of $\xi(\epsilon)$. The steep increase of $\xi(\epsilon)$
after its minimum is a result of our approximations and therefore
unphysical. Notice that the minimum of $\Lambda\xi(\epsilon)<2\pi$.
The initial values for this figure are $\rs/k_{B}T=0.06$ and $t/\Lambda^{2}k_{B}T=0.01$. }

\label{figxiepsilon} 
\end{figure}

To determine the scale at which we may stop integrating the flow equations
and use mean field theory, one may use the scale-dependent correlation
length (see Eq.~\eqref{eq1loopcorr}) $\xi(\epsilon)=\sqrt{\rs(\epsilon)/t(\epsilon)}$.
Mean field theory applies if $\Lambda\xi(\epsilon)\ll2\pi$, while
it is inapplicable otherwise. Thus, the appropriate $\epsilon=\epsilon_{f}$
at which integration may be stopped may be located by integrating
up to the point at which $\xi(\epsilon)$ is a minimum; the mean-field
approach of Sec.~\ref{secdeterminationofQc} may then be used to
determine $Q_{c}$. This is justified \emph{a posteriori} if indeed
$\Lambda\xi(\epsilon_{f})<2\pi$. This condition is satisfied for
all parameter values we have studied (see Fig.~\ref{figxiepsilon}
for a representative example).

\begin{figure}[t]
 \centering \includegraphics[width=8cm]{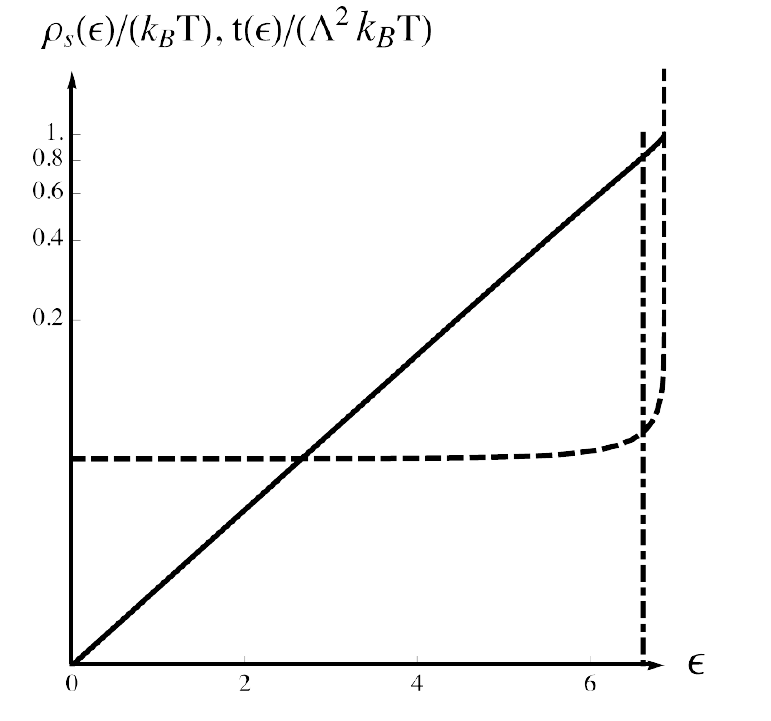} 

\caption{
Demonstration of the asymptotic simple scaling of the trajectories,
demonstrating their scheme independence (see text). The full line is
$t(\epsilon)/\Lambda^2 k_B T$, the dashed line is $\rs(\epsilon)/ k_B T$
and the dot-dashed line indicates the position of the minimum of
$\xi(\epsilon)$ (see Fig.~\ref{figxiepsilon}). The steep increase of
$\xi(\epsilon)$ after its minimum is a result of our approximations and
therefore unphysical (see text). The initial values for this figure are the same as for Fig.~\ref{figxiepsilon}: $\rs/k_{B}T=0.06$ and $t/\Lambda^{2}k_{B}T=0.01$. 
}

\label{figlogplotsflows} 
\end{figure}

How would the calculation described in Sec.~\ref{secdeterminationofQc}
be affected by the RG transformation? As we saw in Sec.~\ref{secdervfloweqns},
$H_{Q}$ is invariant under the RG transformation, while $H_{sG}$
is not, \ie the parameters in $H_{sG}$ will change to $\rs(\epsilon)$
and $t(\epsilon)$, respectively. To obtain the critical $Q_{c}$,
we notice that while $H_{Q}$ is unaffected by the RG, the energy
of a single soliton now depends on $\epsilon$: we have $E_{\mathrm{sol}}(\epsilon)=8\sqrt{\rs(\epsilon)t(\epsilon)}$.
This results in an energy difference between the phase with no solitons
and the phase with a single soliton given by $\Delta E=E_{\mathrm{sol}}(\epsilon)-2\pi\rs(0)Q$.
Setting this equal to zero and solving for $Q$, as in section~\ref{secdeterminationofQc},
yields \begin{equation}
Q_{c}(\epsilon)=\exp(-\epsilon)\frac{4}{\pi}\frac{1}{\rs(0)}\sqrt{\rs(\epsilon)t(\epsilon)}.\label{eqcriticalQcepsilon}\end{equation}
The factor $\exp(-\epsilon)$ ensures that we are using physical (as
opposed to rescaled) units. Eq.~\eqref{eqcriticalQcepsilon} reduces to
the correct mean-field expression, Eq.~\eqref{eqcriticalQc}, for
$\epsilon=0$.

In Fig.~\ref{figxiepsilon} we show a representative plot of the
evolution of $\Lambda\xi(\epsilon)$ with $\epsilon$. Evidently, the
minimum is well below $2\pi$ so that mean field theory is applicable to
the renormalized $\Gamma$.

Let us now discuss the scheme-dependence of our calculation. In Ref.~\onlinecite{schemedependence}
it is shown that the trajectories resulting from the effective action
functional RG scheme we use are scheme-independent, provided that
the quantities $t$ and $\rs$ flow as powers of the parameter $\epsilon$,
\ie $\rs\sim\exp(d_{\rho}\epsilon)$ with some $d_{\rho}$ (and similarly
for $t$). They term the region in which this occurs the {}``freezing
region''. Note that this scheme-independence does not hold in general
for the case of Wilson-type renormalization.

Fig.~\ref{figlogplotsflows} is a log plot of the evolution of $\rs$
and $t$ with $\epsilon$. The full line is $t(\epsilon)$, the dashed
line is $\rs(\epsilon)$ and the dot-dashed line indicates the position
of the minimum of $\xi(\epsilon)$ (see also Fig.~\ref{figxiepsilon}).
Evidently, both $\rs(\epsilon)$ and $t(\epsilon)$ are in the freezing
region at the values of $\epsilon$ that we are interested in. This
happens for all initial values of $\rs$ and $t$ that we have checked.

\begin{figure}[t]
\centering \includegraphics[width=8cm]{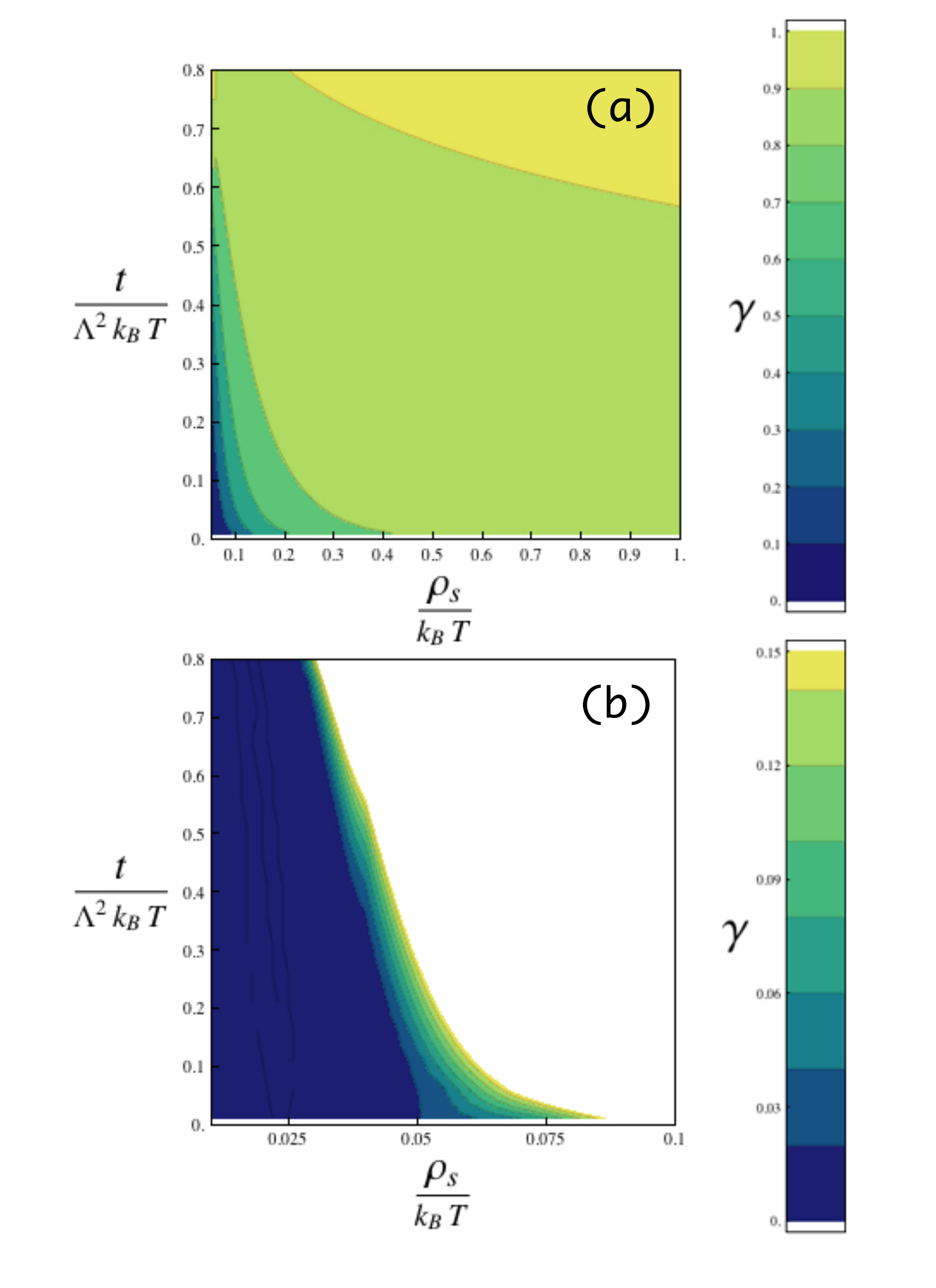} 

\caption{
(Color online) The ratio $\ratio=Q_{c}^{RG}/Q_{c}^{MF}$ predicted by our
RG analysis to its mean-field value. Figure (b) corresponds to
the lower left corner of Fig. (a). The strips on the right indicate
meaning of shade. Notice the difference in scales between the two plots.
White corresponds to $\gamma>0.15$. }

\label{figrslts} 
\end{figure}

\subsection{Results and discussion}

In Figs.~\ref{figrslts} (a) and (b) we show the ratio
$\ratio=Q_{c}(\epsilon)/Q_{c}(0)$ determined by numerically integrating the
flow equations and applying the method discussed above; this corresponds
to the ratio $\ratio=Q_{c}^{RG}/Q_c^{MF}$. A darker colour indicates
a larger decrease of the critical $Q_{c}$ due to thermal effects (see
the insets and notice the different scales). The plot in
Fig.~\ref{figrslts} (b) is a zoomed-in part of Fig.~\ref{figrslts} (a) (notice the range of the axes and also the changed color coding). 

For the purposes of this section, we will switch back to using natural
units, by defining $\tilde{\rs}/k_{B}T=\rs$ and $\tilde{t}/(k_{B}T\Lambda^{2})=t$,
\ie $\tilde{\rs}$ and $\tilde{t}$ are the parameters in physical
units (while $t$ and $\rs$ are in units in which $k_{B}T=\Lambda=1$).

First let us fix $\tilde{\rs}$ and $T$ and vary $\tilde{t}$, \ie
fix $\rs$ and vary $t$. For increasing (decreasing) $t$, $\gamma$
increases (decreases), vanishing as $t\rightarrow0$. This occurs
because, as described in Sec.~\ref{secmodel}, for vanishing $t$
there is no long-range order even in the absence of $Q$. This effect
is not present at mean-field level, hence  $\gamma$ vanishes
with decreasing $t$.

Next, fix $\tilde{t}$ and $T$ and vary $\rs$. From Fig.~\ref{figrslts},
$\gamma$ increases (decreases) for decreasing (increasing) $\rs$
or $\tilde{\rs}$.

Finally, consider fixing $\tilde{t}$ and $\tilde{\rs}$ and varying
the temperature $T$. This corresponds to fixed $t/\Lambda^{2}\rs$
or, in terms of Fig.~\ref{figrslts}, to moving on rays emanating
from the origin with gradient $t/\rs$. Increasing (decreasing) the
temperature $T$ corresponds to moving inwards (outwards) from the
origin. As $T$ is increased, a $T$ will be reached at which the
system will enter the rough phase at the extreme lower left corner
of the diagram; thus, at high enough $T$, $Q_{c}$ vanishes for all
$\tilde{\rs}$ and $\tilde{t}$. Equivalently, this may be viewed
as a proliferation of solitons because the RG flow there is such that
$t(\epsilon)\rightarrow0$ monotonically as $\epsilon\rightarrow\infty$,
so that we also have $E_{\mathrm{sol}}(\epsilon)\rightarrow0$. This
brings up the intriguing possibility of a purely temperature-driven
commensurate-incommensurate transition in suitable systems. For low enough temperatures, $\gamma\rightarrow1$ as expected.

\section{Conclusions}

\label{secconclusions} We have studied thermal effects on the commensurate-incommensurate
transition point of the PT model using a renormalization-group approach.
Our scheme relies on splitting the PT Hamiltonian into a sine-Gordon
part, $H_{sG}$, and a part depending only on the number of solitons
present, $H_{Q}$. We then derive a functional RG transformation which
acts on the sine-Gordon part while leaving the soliton part invariant.
We are thus able to determine the critical $Q_{c}$ at which the
incommensurate phase eventually becomes stable, taking into account
thermal effects. We find a general lowering of $Q_{c}$ compared to
the mean-field result. Furthermore, there exists a regime in the $\rs$-$t$
plane in which $Q_{c}\rightarrow0$ even for finite $t$. This is
due to the roughening transition, which is of Kosterlitz-Thouless
type.

Since real experiments are performed at finite temperatures, the approach
developed here may be applied to describe the C/IC transition in
several physical systems. An application to the case of a Quantum
Hall bilayer at total filling $\nu_{T}=1$ in the presence of a tilted
magnetic field, motivated by recent experiments,\cite{bilayerexperiment2008}
will be described in a forthcoming publication.\cite{unification}


\section*{Acknowledgements}

We acknowledge financial support from the Netherlands Organization
for Scientific Research (NWO) and thank L. K. Lim and N. Hasselmann for
stimulating discussions. We also thank NORDITA for hospitality during the course of this work.

\appendix

\section{Soliton energy}

\label{appsolitonenergy}

The excess energy of a single soliton may be calculated as follows:
We have $E_{\mathrm{sol}}=H[\overline{\theta}]-H[0]$; the solitonic
profile $\overline{\theta}$ satisfies the extremization condition
$\left.\delta H_{PT}/\delta\theta\right|_{\theta=\overline{\theta}}=0$.
This condition gives an Euler-Lagrange equation, the first integral
of which yields \begin{equation}
\frac{\rs}{2}\left(\frac{\partial\overline{\theta}}{\partial x}\right)^{2}=t\left(1-\cos\overline{\theta}\right).\label{eqfirstintegral}\end{equation}
 Using this to eliminate the dependence of $H_{PT}[\overline{\theta}]$
on the cosine term and changing the variable of integration we obtain
\begin{equation}
E_{\mathrm{sol}}=\rs\int_{0}^{2\pi}d\overline{\theta}\frac{\partial\overline{\theta}}{\partial x}.\label{eqesolprelim}\end{equation}
 Finally, inserting Eq.~\eqref{eqfirstintegral} into Eq.~\eqref{eqesolprelim}
and integrating gives $E_{\mathrm{sol}}=8\sqrt{\rs t}$.

\section{Some trigonometric identities}

\label{apptrigident} 

In this Appendix we display some of the calculations leading to Eqs.~\eqref{eqflowtrestricted} and~\eqref{eqflowrsrestricted}.

\begin{widetext} 
It will be useful to know that 
\begin{equation}
\cos^{m}\left(\phi\right)=\begin{cases}
\frac{1}{2^{m-1}}\binom{m}{(m-1)/2}\cos\left(\phi\right)+\ldots, & n\text{ odd}\\
\frac{1}{2^{m}}\binom{m}{m/2}+\frac{1}{2^{m-2}}\binom{m}{(m-1)/2}\cos\left(2\phi\right)+\ldots & n\text{ even}\end{cases}\label{eqcosmphi}\end{equation}
where the dots indicate higher-frequency Fourier components.

We begin by expanding the integrand of Eq.~\eqref{eqflowVgeneral} for $b=1$,
\begin{equation}\begin{split}
	\frac{1}{2}\int \frac{d^2 q}{(2\pi)^2} \frac{1}{\rs q^2+\Lambda^2+V^{(2)}}
	&=
	\frac{1}{2}\int \frac{d^2 q}{(2\pi)^2} 
	\sum_{n=1}^\infty \frac{\left(-V^{(2)}\right)^n}{\left(\rs q^2+\Lambda^2\right)^{n+1}}\\
	&= \frac{1}{4\pi\rs}\sum_{n=1}^\infty \frac{1}{n}\left(\frac{-V^{(2)}}{\Lambda^2}\right)^n \\
	&=-\frac{1}{4\pi\rs} \ln \left(1+ \frac{V^{(2)}}{\Lambda^2}\right)
\end{split}\label{eqsumlog}\end{equation}
\end{widetext}

Next, we find that
\begin{equation}\begin{split}
\ln\left[1+z\cos\left(\phi\right)\right]&
=\sum_{r=1}^{\infty}\frac{z^{r}}{r}\cos^{r}\left(\phi\right)\\
&=\frac{2}{z}\left(1-\sqrt{1-z^{2}}\right)\cos\left(\phi\right)+\ldots\label{eqlogfourierexpn}
\end{split}\end{equation}
 where $\ldots$ indicates higher Fourier components. Eqs.~\eqref{eqsumlog} and~\eqref{eqlogfourierexpn}

Finally, to restrict the flow of $\rs$ to the field-independent subspace, one needs the result 
\begin{equation}
	\frac{1}{2\pi}\int_0^\infty d\phi\,
	\frac
	{\sin^2(\phi)}
	{\left(1+z\cos(\phi)\right)^3}
	=
	\frac{1}{2}\frac{1}{\left(1-z^2\right)^{3/2}}.
	\label{eqprojectrs}
\end{equation}

\end{document}